\documentclass[12pt]{iopart}
\usepackage{iopams}  
\usepackage[all,cmtip]{xy}

\newcommand{\sfq}{\mathsf{Q}}
\newcommand{\sfp}{\mathsf{P}}

\newcommand{\mB}{\mathcal{B}} 
\newcommand{\mE}{\mathcal{E}}
\newcommand{\mF}{\mathcal{F}}

\newcommand{\mS}{\mathcal{S}}

\newcommand{\vn}{\mathbf{n}} 
\newcommand{\vsigma}{\boldsymbol{\sigma}}

\newcommand{\va}{\mathbf{a}} 
\newcommand{\vb}{\mathbf{b}} 

\begin{document}

\title[Quantum uncertainty]
{Quantum Mechanics as a Framework for Dealing with Uncertainty}

\author{Paul Busch}

\address{Department of Mathematics, University of York, York YO10 5DD, UK}
\ead{pb516@york.ac.uk}
\begin{abstract}
Quantum uncertainty is described here in two guises: indeterminacy with its concomitant
indeterminism of measurement outcomes, and fuzziness, or unsharpness. Both features
were long seen as obstructions of experimental possibilities that were available in the
realm of classical physics. The birth of quantum information science was due to the realization
that such obstructions can be turned into powerful resources. Here we review how the utilization
of quantum fuzziness makes room for a notion of approximate joint measurement of noncommuting
observables. We also show how from a classical perspective quantum uncertainty is due
to a limitation of measurability reflected in a fuzzy event structure -- all quantum events are fundamentally
unsharp.

\end{abstract}


\submitto{\PS}
\maketitle

\section{Introduction}

Quantum physics arrived with a bang: the realization that experimental research 
had reached the limits of validity of classical physics. It was an understandable
shock reaction of the pioneers of quantum mechanics to  see this new theory as 
encapsulating obstructions  -- to the observation of atomic phenomena, to 
the application of the concepts of classical physics to atomic objects.

Thus, for example, the complementarity and uncertainty principles were commonly regarded as
expressions of limitations: the former states the necessity of applying both
the classical particle picture and the classical wave picture in the description
of microsystems as well as the impossibility of the simultaneous application
of these pictures; the latter was taken as quantifying this restriction of the simultaneous definition 
and measurement of canonically conjugate pairs of variables.

Another, related sort of limitation or obstruction was seen in the irreducibly probabilistic nature
of quantum mechanics, which reflects the inherent indeterminacy of the values of observables
and the fundamental indeterminism in the occurrence of measurement outcomes. This 
observation led directly to the quest for a crypto-deterministic, hidden variable description 
supposedly underlying quantum mechanics.  The well-known no-hidden-variables theorems
of von Neumann, Bell, Kochen and Specker, and others describe the extent to which
any such description (as, for example, the de Broglie-Bohm theory) must differ from a 
classical-physical description in order to be consistent with quantum mechanics.

Entanglement is another fundamental quantum feature that was identified as leading to
unpleasantly strange nonclassical behaviour -- quantum nonlocality, as was noted in 1935 by 
Einstein, Podolsky and Rosen and by Schr\"odinger. Entanglement entails a limitation in the 
definition of the state of an individual object independently of its environment. 

While quantum mechanics was enormously successfully applied in the analysis and exploitation 
of many new physical phenomena, it took many decades during which the above foundational issues 
were revisited and reviewed from new angles until finally the realization dawned that rather than
being obstacles they are potential resources for information processing protocols which, if feasible, 
would be greatly more powerful in principle than methods based on classical physics. Hence
we are witnessing a change of perspective from seeing quantum structures as {\em obstructions} to
exploring them as {\em resources}. 

In this contribution I will briefly review the notion of {\em quantum uncertainty} in its two 
guises as  {\em indeterminacy} and {\em unsharpness}, and  I will indicate how quantum uncertainty 
is fundamental to the interplay of quantum obstructions and novel quantum resources referred to above (Sec. \ref{sec:qu}).
This is complemented with a sketch of a consistent and comprehensive way of presenting quantum mechanics 
as a classical fuzzy probability theory which makes precise the sense in which {\em from a classical
perspective}, this theory reflects limitations of measurability. The existence of this particular classical
representation of quantum probability also demonstrates that the two forms of quantum uncertainty, 
indeterminacy and fuzziness, are in a way interchangeable and can be traded for each other (Sec. \ref{sec:qu-class}). 

I dedicate this paper to Pekka Lahti on the occasion of his 60th birthday -- we have spent many years trying together to 
understand aspects of the fundamental quantum features discussed here.

\section{Quantum obstructions, or things to do with quantum uncertainty}\label{sec:qu}

\subsection{Quantum uncertainty as indeterminacy}

In a letter to Max Born dated December 4, 1926, Albert Einstein wrote these famous words:

\begin{quote}{\small
Quantum mechanics is very worthy of regard. But an inner voice tells me that this is not yet the right track. 
The theory yields much, but it hardly brings us closer to the Old oneÕs secrets. I, in any case, am convinced
that He does not play dice.}
\end{quote}
This is Einstein's rejection of the conclusion, suggested by Born's probabilistic interpretation  of quantum 
mechanics \cite{Born26},  that the world should be fundamentally indeterministic. Instead of accepting
{\em quantum uncertainty}, he initiates the search for {\em hidden variables}, a possibility hinted at by Born 
in his paper. 

In classical probabilistic physical theories, all quantities are assigned sharp, definite values in every pure state.
Quantum uncertainty in its first guise is the statement that within quantum mechanics, there is no state in which
all observables would have definite values; and for every observable there are pure states (namely, superpositions
of its eigenstates) in which their values are not definite, that is, {\em indeterminate}. If a state is understood as a 
probability catalogue for all measurement outcomes, then an observable would have a definite value in a given 
state if the probability distribution for the values of that observable is 0-1-valued. Gleason's theorem asserts that 
all states, defined as probability measures over the set of all closed subspaces of the complex  Hilbert space (of
dimension greater than 2)
associated with a quantum system, are given by some density operator $\rho$ via the trace formula:
\begin{equation}
prob(P)=trace(\rho\, P),
\end{equation}
where $P$ is the orthogonal projection onto a closed subspace and could figure as a spectral projection of
a self-adjoint operator representing an observable. It is evident that quantum states are {\em not} 0-1-valued
(or dispersion-free). 

Gleason's theorem thus presents a severe obstacle for any attempt to supplement 
quantum mechanics with a classical description. Subsequent studies have shown that any hidden 
variable theory that reproduces all empirical predictions of 
quantum mechanics must incorporate some form of contextuality, that is, a dependence of the probability 
assignments on the measurement or preparation context. This is known to be true, in particular, for the best 
established hidden variable approach due to de Broglie and Bohm. Rather than recalling the formal ingredients
of a hidden variable theory in general, we will point these out in the construction of a classical embedding of quantum
probability theory in the next section.

The Kochen-Specker and Bell theorems are refinements of Gleason's theorem, giving rise to an
industry of attempts to specify minimally small sets of quantum propositions (projections) on which 
a consistent assignment  of values 1 or 0 (true or false) cannot be defined (for an introduction and 
survey of this ``coloring problem",  see \cite{QTCM}). More recently, the converse problem has been 
solved, of describing large sets (of appropriate proposition structures) of projections that can be 
consistently assigned values 0 or 1. The result is laid down in the Bub-Clifton uniqueness theorem 
for interpretations of quantum mechanics \cite{IQW}. Each possible  rule of defining subsets of 
projections with definite values involves the choice of a quantum state and a reference observable, 
and can thus be taken as defining an interpretation of quantum mechanics by way of fixing the notion 
of reality of properties. The standard eigenstate-eigenvalue rule and Bohm's interpretation with a definite 
position variable are found to be examples of this general scheme. In this way one has learned to handle 
and quantify the limitation of ascribing truth values to quantum physical propositions.
 
Quantum indeterminacy is closely related the impossibility of distinguishing all pairs of distinct states through a 
single-shot measurement. In fact, suppose two different states $\rho_1$ and $\rho_2$ could be distinguished
by the measurement of a single observable (represented in general by a positive operator measure, POM).
There would thus be a pair of complementary outcomes, represented by positive operators (effects) 
$E_1$ and $E_2$ (such that $E_1+E_2=I$, the identity operator), such that $\rho_1$ will always trigger 
the outcome associated with $E_1$ and never that associated with $E_2$, and similarly $\rho_2$ will always
lead to $E_2$ and never to $E_1$. It follows that  the ranges of $\rho_1,\rho_2$ are mutually orthogonal
subspaces (in fact, contained in the closed eigenspaces of $E_1,E_2$ associated with the eigenvalues 1, 
respectively. Hence nonorthogonal pairs of distinct states cannot be distinguished by a single measurement.

This impossibility, or quantum obstruction, has turned out to be fundamental to the security of quantum cryptographic 
protocols as it makes eavesdropping by way of single shot measurements impossible. It also makes superluminal
signaling impossible.

A pure state of a compound system is {\em entangled} if it is not {\em separable}, i.e., if it is not a product state. 
This is equivalent to saying that there is no observable with nondegenerate eigenstates of product form whose 
values are definite. In this sense entanglement is an instance of quantum indeterminacy.

\subsection{Quantum indeterminacy and indeterminism}

The above Einstein quote alludes to the probabilistic nature of quantum mechanics, hence the 
{\em indeterminism} of this theory, which Einstein saw as a deficiency to be remedied in the course 
of future developments. We referred to the connection with the probability structure and the impossibility 
of defining noncontextual truth value assignments to all experimental propositions of a quantum system, 
and arrived at the conclusion that according to quantum mechanics there is a fundamental {\em indeterminacy}
of the values of most observables in any given state.

 If the values of physical quantities are not definite, it would
seem natural to conclude that a measurement of that quantity would not yield a predictable outcome. Indeed,
it is commonly accepted that quantum mechanics only provides the probabilities for these outcomes.
Hidden-variable theories are designed to restore the definiteness of (at least) some quantities, the values of which
would then determine the outcomes of measurements. Thus determinateness should restore determinism.
However, the necessary ``hidden-ness" of these variables is just a representation of the fact that these hidden
values cannot be accessed (measured or known) {\em as a matter of principle}. It seems a question of semantics
whether or not one takes this observation as justification for the continued use of the term indeterminacy. 

According to quantum mechanics, there is no empirically accessible cause for the occurrence of a particular 
measurement outcome if the state is not an eigenstate of the measured observable. The randomness of the
outcomes is grounded in the fundamental indeterminateness and indeterminism that we call quantum uncertainty.

This uncertainty of measurement outcomes for quantum objects is being exploited in the theory of {\em quantum games}
which often show peculiar advantages for a party using quantum strategies compared to classical games.

\subsection{Quantum uncertainty as fuzziness}

Nonorthogonal pure states cannot be distinguished by a single-shot measurement. Geometrically, such states
belong to different orthonormal bases, and as such they are eigenstates of {\em noncommuting observables}.
Thus we see that quantum uncertainty is closely related to the existence of incompatible pairs of observables.
The above proof of the indistinguishability of nonorthogonal states makes implicit use of the fact that the two 
observables of which these states are eigenstates cannot be measured together. 

The Heisenberg uncertainty principle comprises three physical statements which actually have been proven
as theorems in quantum mechanics for certain pairs of observables; we phrase them here informally: 
\begin{itemize}
\item[(a)] the values of two noncommuting 
quantities can be {\em unsharply defined} to the extent allowed by the uncertainty relation for the widths of their 
distribution in the given state; 
\item[(b)] the values of two noncommuting quantities can be {\em jointly and approximately measured} to accuracies 
allowed by a measurement uncertainty relation;
\item[(c)] the initially sharp value of a quantity $A$ will be disturbed (made unsharp) by a subsequent 
measurement of a noncommuting quantity $B$ such that the inaccuracy of the $B$-measurement and the 
magnitude of the disturbance of $A$ obey an uncertainty relation.
\end{itemize}
These are the state preparation, joint measurement, and disturbance versions of the uncertainty relation.
The last version is an expression of the {\em Heisenberg effect}. 

Note that the above formulations state {\em positive} possibilities, in contrast to the traditional way of phrasing
the uncertainty principle as a {\em limitation} of preparations or measurements. Statement (a) is well known and
universally accepted whereas (b) and (c) have remained contentious for many decades, due to the lack of 
a rigorous formulation. These latter two statements
have been made precise only rather recently on the basis of the generalised representation of observables as POMs,
which allowed the introduction of operationally relevant concepts of approximate (joint) measurements and of 
suitable measures of inaccuracy.

\subsubsection{Joint measurability.}
Two observables, represented as POMs $M_1,M_2$ on the real line $\mathbb{R}$ (say), are jointly 
measurable if they are marginals of a third observable $M$ defined on $\mathbb{R}^2$; this means that
for all (Borel) subsets $X,Y$ of $\mathbb{R}$, one has $M_1(X)=M(X\times\mathbb{R})$ and 
$M_2(Y)=M(\mathbb{R}\times Y)$. This notion captures the idea that if two quantities can be measured together
they must have a joint probability distribution for every state.

It is well known that two {\em sharp observables} (represented by spectral measures) are jointly measurable if and only
if they commute, but two noncommuting {\em unsharp observables} (POMs that are not projection valued) can be
jointly measurable. Unsharpness is thus a prerequisite for joint measurability.

There are not many general results on the joint measurability of pairs of noncommuting observables. 
Early positive observations after von Neumann's and Wigner's no-go theorems (the latter based on
the failure of the Wigner function to be nonnegative) were the discovery of the Husimi function, which was
much later understood to give rise to an instance of a covariant phase space observables. It became gradually
clear that suitable smeared versions of position and momentum are indeed jointly measurable, and that in
all instances an inaccuracy tradeoff relation is satisfied. This led to a notion of unsharp observables and
approximate (joint) measurements which can be expected to provide the basis for a general theory of joint 
measurability in quantum mechanics (Fig. 1).
\begin{figure}
$$
\xymatrix{
& G(X\times Y)\ar[ldd]_{\tiny [Y=\Bbb R]}\ar[rdd]^{\tiny [X=\Bbb R]} &\\
&&\\
G_1(X)\ar@{~>}[d]&&G_2(Y)\ar@{~>}[d]\\
E(X)&&F(Y)}
$$
\caption{Scheme of an approximate joint measurement of two (sharp or unsharp, noncommuting) observables
$E,F$ with values in $\mathbb{R}$. The measured observable $G$ has values in $\mathbb{R}^2$, its Cartesian 
marginals $G_1,G_2$ are approximators to $E,F$, respectively. The quality of approximation can be 
quantified by means of suitable measures of (in)accuracy.}
\end{figure}
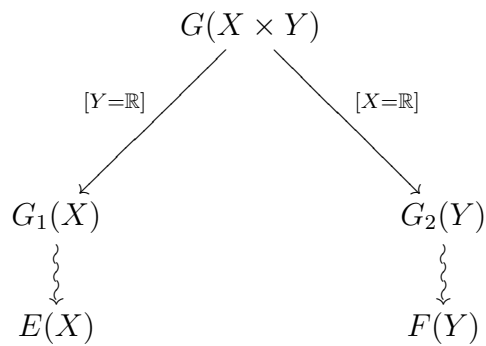
 
The question of the approximate joint measurability of position and momentum observables and that of 
pairs of qubit observables have been analysed in great generality and are now well understood. 
The solution to the joint measurement problem for position-momentum case was obtained in a breakthrough 
paper by Werner \cite{Werner04b}. It is now known that a trade-off relation 
\begin{equation}\label{eqn:m-ur}
d(Q,M_1)\,d(P,M_2)\ge C\hbar\qquad  (C>0)
\end{equation}
must hold for measures of the distances $d(Q,M_1)$ and $d(P,M_2)$ between position
and momentum on the one hand and a pair of approximating observables (POMs) $M_1$ and 
$M_2$ on the other hand if the latter are to be jointly measurable. This is, finally, Heisenberg's famous though debated
{\em joint measurement inaccuracy relation} made rigorous in operationally meaningful terms. Alternative formulations
of (\ref{eqn:m-ur}) based on a measure of {\em error bar widths} have been found subsequently \cite{BuPe06}.
It turns out that finite Werner distance implies finite error bar widths and that there are more jointly measurable
pairs of observables $M_1,M_2$ approximating $Q,P$ in the sense of finite error bar widths than there are with
finite Werner distances. For a recent review of Heisenberg's uncertainty principle exemplified for position and 
momentum, see \cite{BuHeLa06}.

Approximate joint measurements of qubit observables were studied in \cite{BuHe07} and the joint 
measurability of a pair of simple qubit observables was characterised independently in 
\cite{BuSchm08,StReHe08,Liu-etal08}. We recall this latter result briefly, starting with the example of a POM
that represents a joint measurement of {\em all} spin directions of a spin-$\frac 12$ system.

Let $S^2$ denote the unit sphere, with its $\sigma$-algebra of Borel sets ${\cal B}(S^2)$ and the
rotationally invariant measure $d\Omega(\vn)$, normalized as $\Omega(S^2)=4\pi$. (Here $\vn$ denotes
a unit vector labeling a point on $S^2$.) Then the following defines a normalized POM:
\begin{equation}
{\cal B}(S^2)\ni Z\mapsto G(Z):=\frac 1{2\pi}\int_Z \frac12(I+\vn\cdot\vsigma)\,d\Omega(\vn)\,.
\end{equation}
Now, if $Z( \pm\vn_o)$ denotes the hemisphere with centre $\pm\vn_o$, we obtain:
\begin{equation}
G\left(Z(\pm\vn_o)\right)=\frac 12(I\pm \frac 12\vn_o\cdot\vsigma)\,.
\end{equation}
Note that $G\left(Z(\vn_o)\right)+G\left(Z(-\vn_o)\right)=I$, so that these two positive operators constitute a 2-valued
(so-called simple) observable. Each of these observables are smeared or fuzzy versions of the associated
sharp observables defined by the projections $P(\pm\vn_o):=\frac 12(I\pm\vn_o\cdot\vsigma)$, in the sense that one can write
$\frac 12(I+\frac 12\vn_o\cdot\vsigma)=\frac 34P(\vn_o)+\frac 14P(-\vn_o)$ and similarly for the other effect.
 One can also say that these
smeared versions are approximations to the corresponding sharp observables. Thus the observable $G$ represents 
an approximate joint measurement of all sharp spin projections.

In general, a qubit effect can be represented as an operator $A=a_0I+\va\cdot\vsigma$, with the constraint that
the eigenvalues lie in $[0,1]$, hence $0\le a_0\pm|\va|\le 1$, or:
\begin{equation}
|\va|\le\min(a_0,1-a_0).
\end{equation}
The result mentioned above now reads as follows. Two qubit effects
$A=a_0I+\va\cdot\vsigma$, $B=b_0I+\vb\cdot\vsigma$ are jointly measurable 
if and only if they satisfy the inequality
\begin{equation}\label{eqn:coex}
\frac 12[\mF(2-\mB)+\mB(2-\mF)]+(xy-4\va\cdot\vb)^2\ge 1.
\end{equation}
Here the following abbreviations are used:
\begin{eqnarray*}
\mF&:=&\varphi(A)^2+\varphi(B)^2;\\
 \mB&:=&\beta(A)^2+\beta(B)^2;\\
x&:=&\varphi(A)\beta(A)=2a_0-1;\label{eqn:x}\\
y&:=&\varphi(B)\beta(B)=2b_0-1;\\
\varphi(A)&:=&\sqrt{a_0^2-|\va|^2}+\sqrt{(1-a_0)^2-|\va|^2};\label{eqn:fa2}\\
\beta(A)&:=&\sqrt{a_0^2-|\va|^2}-\sqrt{(1-a_0)^2-|\va|^2}.\label{eqn:ba2}
\end{eqnarray*}
The quantities $\varphi(A)$ and $\beta(A)$ are measures of 
{\em unsharpness} (or fuzziness) and {\em bias}, respectively \cite{BuSchm08,Busch09-sharpness}.
In the
unbiased case, where $a_0=b_0=\frac 12$, Eq. (\ref{eqn:coex}) assumes the much simplified form
\begin{equation}
16 |\va\times\vb|^2\le (1-4|\va|^2)(1-4|\vb|^2)\,,
\end{equation}
or equivalently,
\begin{equation}
|\va+\vb|+|\va-\vb|\le 1\,.
\end{equation}
The two factors on the right hand side of the former inequality are also measures of the unsharpness
of the two effects $A,B$ and are thus seen to be constrained by the noncommutativity of $A,B$
(as quantified by the vector product term on the left hand side, which is proportional to $\|[A,B]\|^2$).

With the choices $\va=\frac 14\vn_o$, $\vb=\frac 14\vn_o'$, we reproduce the effects $G\left(Z(\vn_o)\right)$, 
$G\left(Z(\vn_o')\right)$, and it is easily seen that the last two inequalities are satisfied in this case.
This inequality describes the degree of unsharpness in the two noncommuting effects required for them
to be jointly measurable.

We conclude that the strict obstruction to the joint measurability of noncommuting quantities can be lifted
if a sufficient degree of fuzziness is allowed in the definition of the observables and their measurements.
This step opens the door to the introduction of {\em informationally complete} observables, whose statistics
allow the complete identification of all states and thus give rise to realizations of quantum state tomography
protocols.

This transformation of quantum uncertainty from vice to virtue was a move envisaged by Heisenberg in 1927
who, however,  lacked the formal tools to make precise the corresponding measurement versions of his uncertainty 
principle. In his famous Como lecture \cite{Bohr28}, Bohr endorsed this positive outlook as follows:
\begin{quote}
In the language of the relativity theory, the content of the relations (2) [the
uncertainty relations] may be summarized in the statement that according to the
quantum theory a general reciprocal relation exists between the maximum
sharpness of definition of the space-time and energy-momentum vectors
associated with the individuals. This circumstance may be regarded as a simple
symbolical expression for the complementary nature of the space-time
description and claims of causality. At the same time, however, the general
character of this relation makes it possible to a certain extent to reconcile
the conservation laws with the space-time co-ordination of observations, the
idea of a coincidence of well-defined events in a space-time point being
replaced by that of unsharply defined individuals within finite space-time
regions.
\end{quote}
Incidentally, this passage appears to be the first occurrence of the word {\em unsharp} 
in the quantum physics literature; hence it would seem that this teutonic extension to the English vocabulary
is due to Bohr.

\subsubsection{Heisenberg effect.}
Having introduced the idea  of the approximation of one observable by another observable, 
one can use the associated measures of distance or inaccuracy to quantify the inevitable disturbance 
of a quantum state through a measurement. If the system is in an eigenstate of the measured quantity,
then a L\"uders measurement of that quantity does not alter the state (see, e.g., \cite{QTM}). However,
consider a measurement of position on a state in which the momentum is fairly well defined (as represented
by a rather sharply peaked distribution). If the position measurement is of good quality, the momentum
distribution afterwards will no longer be sharply peaked. 

This intuitive consideration can be made precise: if the position measurement is followed by a momentum
measurement, one will be able to see the disturbance of the momentum through the position measurement:
the final momentum measurement statistics can be expressed in terms of the state prior to the interfering
position measurement, and this leads to the definition of a POM  $M$ that would be the sharp momentum
observable if no position measurement had been made. If the position measurement is sharp, then it can
be shown that the observable $M$ must actually commute with the position observable and thus gives
no information at all about the momentum distribution of the initial state (Fig.\ 2(a))! 
If instead of sharp position an
approximate  measurement of position is made in an appropriate way, then the subsequent momentum
measurement defines an unsharp momentum observable $M$ relative to the initial state (Fig.\ 2(b)). 
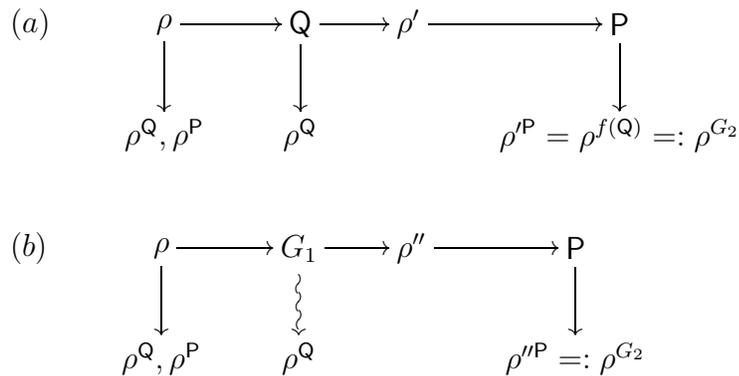
\begin{figure}
$$
\leqno{(a)} \qquad\xymatrix{
\rho\ar[r]\ar[d]&{\sfq}\ar[r]\ar@{->}[d]&\rho'\ar[r]&{\sfp}\ar@{->}[d]\\
\rho^{\sfq},\rho^{\sfp}&\rho^{\sfq}&&\rho'^{\sfp}=\rho^{f(\sfq)}=:\rho^{G_2}
}
$$
$$
\leqno{(b)}\qquad\xymatrix{
\rho\ar[r]\ar[d]&{G_1}\ar[r]\ar@{~>}[d]&\rho''\ar[r]&{\sfp}\ar@{->}[d]\\
\rho^{\sfq},\rho^{\sfp}&\rho^{\sfq} &&\rho''^{\sfp}=:\rho^{G_2}
}
$$
\caption{Momentum disturbance through position measurement. (a) A (nonselective) measurement of 
sharp position, $G_1:=\sfq$,  leads to a state $\rho'$, on which a measurement of sharp momentum is performed. This
defines a POM $G_2$ relative to $\rho$ whose effects are functions of position and thus $G_2$ is not even an approximation to
the sharp momentum $\sfp$. (b) A suitable form of measurement of an approximate position observable $G_1$ is performed,
leaving the system in a state $\rho''$, on which a measurement of sharp momentum is carried out. This defines a POM
$G_2$ relative to $\rho$ which is an approximation to $\sfp$.}
\end{figure}
In fact, the 
whole sequence of both measurements constitutes a joint approximate measurement of position and 
momentum relative to the initial state. (Details of the proof are reviewed in \cite{BuHeLa06}.) 
The degree of inaccuracy of the resulting approximate momentum measurement represented by $M$
is a measure of the disturbance of the original, undisturbed (by the position measurement) momentum 
distribution, and the inaccuracy  inequality (\ref{eqn:m-ur}), which holds for this joint (sequential) 
measurement, describes the tradeoff between the position measurement inaccuracy and the disturbance of
momentum, thus confirming version (c) of the uncertainty principle.

The Heisenberg effect -- the necessary disturbance of the quantum state through measurement, can thus be
quantified by suitable joint measurement inaccuracy relations. If an eavesdropper observes a quantum 
communication channel, her measurement will disturb the transmitted qubit unless the state is an eigenstate
of the chosen measurement, but this is beyond the control of the eavesdropper and thus the measurement 
leaves a detectable trace. In this way the Heisenberg effect becomes decisive for the security of cryptographic
protocols.

We conclude that the noncommutativity of pairs of quantum observables necessitates the introduction of a degree of
fuzziness in order to enable approximate joint measurability of such pairs.

\section{Quantum probability as classical fuzzy probability}\label{sec:qu-class}

In quantum mechanics the distinction between sharp and unsharp observables is operationally meaningful \cite{OQP}.
Still, there is a fundamental degree of fuzziness even in the case of sharp quantum observables, represented by
projection valued measures. Consider, for example, two rank-1 projections $P_1$ and $P_2$ that are neither 
orthogonal nor identical. In the pure state $\rho=P_1$, the probability for $P_2$ is neither 1 nor 0. By contrast, in the
corresponding classical situation this latter probability would be 0. We now show that from a classical perspective, 
{\em all} effects of a quantum system are fuzzy properties, whether they are projections or just non-idempotent positive
operators.

Let $\mS_q$ and $\mS_c$ denote the convex sets of (probabilistic) states of a quantum or classical system, respectively, 
with $\mE_q$ and $\mE_c$ being the associated sets of effects. Thus $\mS_q$ is the set of all density operators, 
$\mE_q$ the set of all positive operators $E$ such that  $O\le E\le I$. Each effect defines a unique affine functional
on the set of states, which is given by the trace formula, 
\[
E[\rho]:=\tr[\rho E]\equiv p_\rho(E). 
\]
This is the probability that the measurement of $E$ give a positive outcome indicating the occurrence of the event associated
with $E$ in state $\rho$.

We can think of $\mS_c$ as (a dense convex subset of)
the convex set of all probability probability measures on a measurable space $(\Omega,\Sigma)$ and $\mE_c$ the 
set of all measurable functions $f:\Omega\to\mathbb{R}$ with values between 0 and 1. Each classical effect is  thus a
fuzzy or crisp set and defines an affine functional on the set of probability measures via 
\[
f[\mu]:=\int_\Omega fd\mu\equiv p_\mu(f),
\]
giving the probability of the event associated with effect $f$ in state $\mu$.

To approach the formalization of a hidden-variable description of quantum mechanics, we note that there 
are two canonical ways of relating the quantum statistical model $(\mS_q,\mE_q)$ with a classical statistical 
model $(\mS_c,\mE_c)$. 

\subsection{Classical embedding}
Let $\Phi:\mS_q\to\mS_c$ be an affine mapping (i.e. a mapping that preserves convex combinations). This fixes
a ``dual" mapping $\Phi':\mE_c\to\mE_q$ via 
\[
p_\rho(\phi'(f))\equiv\Phi'(f)[\rho]=f[\Phi(\rho)]\equiv p_{\phi(\rho)}(f)
\] 
for all $\rho\in\mS_q$, $f\in\mE_c$. 
The map $\Phi'$  is interpreted as a {\em quantization map}.
If one wants to ensure that all quantum effects are covered, that is, $\Phi'$ is surjective, it follows that $\Phi$ must be
injective. There is a unique family of solutions which is easily characterized: each such injective affine map $\Phi$
is generated by (and conversely defines) a unique informationally complete observable 
$A:\Sigma\to\mE_q$ via $\Phi(\rho)=p_\rho^A$, where
\[
p_\rho^A(X)=\tr[\rho A(X)],\quad X\in\Sigma.
\]
Then every quantum state is uniquely associated with a probability measure; but the desired surjectivity of $\Phi'$ has
not been achieved: it can be shown not every quantum effect is an image of a classical effect. Only a suitably defined
(unique) linear extension of $\Phi'$ to all bounded measurable functions is surjective. This means that some quantum
effects are represented by functions $f$ that are not nonnegative. This deficiency is reminiscent of the ``dual" deficiency
of the Wigner function representation of quantum states.

We conclude that classical embeddings via informationally complete observables give rise only to {\em partial}
classical representations of a quantum statistical model. Still, this representation is almost complete in a 
formal sense as all the effects in the range of the observable $A$ are in the range of $\Phi'$ (in fact, $A(X)=\Phi'(1_X)$,
where $1_X$ denotes the indicator function of the set $X$), and their span is dense 
in the space of selfadjoint bounded operators.

\subsection{Classical extensions}
The only way of defining a comprehensive classical representation of a quantum statistical model via 
an affine correspondence is through a {\em reduction map} $\Psi:\mS_c\to\mS_q$ and its associated dual map
$\Psi':\mE_q\to\mE_c$. Since one wants coverage of all quantum states, $\Psi$ is required to be surjective. 
A solution was introduced by Misra \cite{Misra74} and cast in the framework of quantum and classical statistical dualities 
by Bugajski \cite{Bug91,Bug93a}. (A detailed review of the literature on this subject is found in \cite{BuSt08}.)

Let now $\Omega_q:=\mS_q^{pure}$ be the set of pure quantum states, represented as rank-1 
projections, equipped with its natural algebra $\Sigma$ of Borel subsets.  (For a qubit this is the 
surface of the Bloch sphere.) Let $\delta_\omega$ denote the point (Dirac) measure 
concentrated on $\omega\in\Omega_q$. Then we define:
\begin{equation}
\mS_c\ni\mu=\int_{\Omega_q}\delta_\omega\,d\mu(\omega)\mapsto\Psi_M(\mu):=
\int_{\Omega_q} \omega\,d\mu(\omega)\equiv\rho_\mu\in\mS_q
\end{equation}
Since every convex decomposition of a density operator $\rho$ can be cast in the form of such an integral,
it is clear that this affine map $\Psi_M$ is surjective.
Noting that 
\[
\tr[\rho_\mu E]=\int_{\Omega_q}\tr[\omega E]d\mu(\omega)=\int_{\Omega_q} f_E(\omega)\,d\mu(\omega)
\]
we obtain the dual correspondence
\[
\mE_q\ni E\mapsto\Psi'_M(E)=f_E\in\mE_c, \qquad f_E(\omega)=\tr[\omega E].
\]
Here it is seen that every quantum effect $E$ -- whether sharp (i.e., projection) or unsharp -- is represented classically
as a fuzzy set $f_E$. The set of all classical effects of this form is a subset of the full set of classical effect and not
sufficient to separate all classical states. 

The classical representation of quantum mechanics through the above map $\Psi_M$ was conceived rather intuitively 
(and in fact formulated rigorously) in the work of Misra. This leaves open the question whether there are any alternative
constructions. It has been shown recently \cite{BuSt08} that the map $\Psi_M$ gives the essentially unique non-redundant
representation in the following sense. Nonredundance means that one starts with a phase space $\Omega$ and 
requires that the reduction map $\Psi$ is such that
there is a correspondence $\iota$ between $\Omega$ and the set of pure quantum states $\Omega_q$, 
$\iota(\omega)=\Psi(\delta_\omega)$. As a reduction map, $\Psi$ also has to satisfy a certain physically natural 
continuity property. It then follows that $\Psi$ can be represented according to
\begin{equation}
\tr[\Psi(\mu) E]=\int_\Omega\tr[\iota(\omega)E]\mu(d\omega)=\int_{\Omega_q}\tr[\omega' E](\mu\circ\iota^{-1})(d\omega'),
\end{equation}
where $\mu\in\mS_c$ and $E\in\mE_q$. This generalized representation $\Psi$ differs from the Misra map $\Psi_M$
by the map $\iota$, which essentially effects a relabeling of the set of pure quantum states 
($\Psi(\mu)=\Psi_M(\mu\circ\iota^{-1})$). The dual map is $\Psi'(E)=f_E$, with $f_E(\omega)=\tr[\iota(\omega)E]$. This
shows that also in this generalized case all quantum effects are represented as classical fuzzy sets. Examples of maps $\Psi$ with different $\iota$ are worked out in \cite{BuSt08}.

\section{Conclusion}

We have reviewed the notion of quantum uncertainty in its two forms of indeterminacy and fuzziness, 
or unsharpness, and we have shown how quantum uncertainty underlies many of the infamous 
quantum restrictions that have recently  been rediscovered as resources for information processing. 
We showed how quantum fuzziness gives room for the approximate joint 
measurability of noncommuting observables, which is necessary for informational completeness and
quantum state tomography. 

We also have reviewed the only ``good" 
classical representation of the quantum statistical model $(\mS_q,\mE_q)$, and found that it confirms 
the known no-go theorems for hidden variable supplementations of quantum mechanics: this representation  
preserves quantum uncertainty in  the form of a fundamental fuzziness.
In fact, the distinction between sharp and unsharp quantum effects has been blurred as all quantum effects are
represented as fuzzy classical effects. Even though the classical statistical model contains all Dirac measures, that is, the pure states, classical events on which they are dispersion free, the sharp classical properties, do not occur in the representation.
One may say that indeterminacy and fuzziness have become indistinguishable.

From this classical perspective, quantum uncertainty is due to a limitation
of measurability: only effects of the form $f_E$ are measurable, and their measurement does not allow one to
distinguish the different possible convex decomposition of a mixed quantum state. Nor is it possible in a single shot
measurement to pinpoint a pure state.

\end{document}